\documentclass[twocolumn,showkeys,aps,prb,showpacs]{revtex4-1}
\usepackage{graphicx}
\usepackage[CJKbookmarks,dvipdfm,colorlinks,linkcolor=blue,citecolor=blue]{hyperref}

\begin{document}

\title{Thermoelectric properties of  orthorhombic group IV-VI monolayers from the first-principles calculations}

\author{San-Dong Guo}
\affiliation{Department of Physics, School of Sciences, China University of Mining and
Technology, Xuzhou 221116, Jiangsu, China}
\begin{abstract}
Two-dimensional (2D) materials  may have potential  applications in thermoelectric devices. In this work, we systematically investigate  the thermoelectric properties of  orthorhombic group IV-VI monolayers $\mathrm{AB}$ (A=Ge and Sn; B=S and Se)
by the  first-principles calculations and semiclassical Boltzmann transport theory.
The spin-orbit coupling (SOC)  is included  to investigate their electronic transport, which produces observable effects on power factor, especially for n-type doping.  According to calculated $ZT$,  the four monolayers exhibit diverse anisotropic thermoelectric  properties, although they have similar hinge-like crystal structure. The GeS along  zigzag and armchair directions shows the strongest anisotropy,  while  SnS and SnSe show  mostly isotropic efficiency of thermoelectric conversion, which can be understood by the strength of anisotropy of their respective power factor, electronic and lattice thermal conductivities.
Calculated results show that $ZT$ for different carriers of n- and p-type  has  little difference for GeS, SnS and SnSe.
It is found that  GeSe, SnS and SnSe  show better thermoelectric performance compared to GeS in n-type doping, and SnS and SnSe  exhibit higher  efficiency of thermoelectric conversion in p-type doping. Compared to  a lot of 2D materials,  orthorhombic group IV-VI monolayers $\mathrm{AB}$ (A=Ge and Sn; B=S and Se) may possess better thermoelectric performance due to higher power factor and lower thermal conductivity. Our work would be beneficial to further experimental study.
\end{abstract}


\maketitle

\section{Introduction}
Thermoelectric materials, which can directly convert  heat to electricity or vice versa  and  make  essential contributions
to energy crisis and global warming, have been a hot spot\cite{s1000,s2000}.
The conversion efficiency of thermoelectric materials can be measured by the dimensionless  figure of merit $ZT=S^2\sigma T/(\kappa_e+\kappa_L)$, in which S, $\sigma$, $T$, $\kappa_e$ and $\kappa_L$ are the Seebeck coefficient, electrical conductivity, absolute  temperature, the electronic and lattice thermal conductivities, respectively.
The high-performance thermoelectric materials  require a high power factor ($S^2\sigma$) and a low thermal conductivity ($\kappa=\kappa_e+\kappa_L$), leading to large $ZT$ value.  Unfortunately, they are generally coupled with each other, and enhancing one  can have the opposite effect on another.  Due to simultaneously increasing  power factor and decreasing thermal conductivity\cite{dw}, low-dimensional materials may have important potential advantages to improve efficiency of thermoelectric conversion, and a lot of  works have focused on 2D materials, such as monolayer phosphorene,  monolayer silicene  and  semiconducting transition-metal dichalcogenide monolayers\cite{s1,s2,s3,s4,s5,s6,s7}.
 \begin{figure}
 \includegraphics[width=8cm]{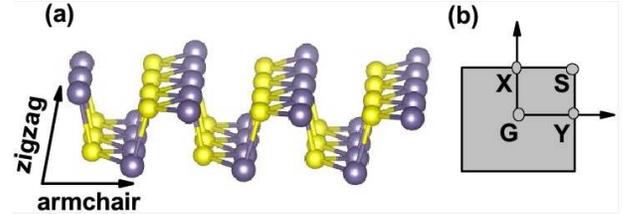}
 \caption{(Color online) (a) The crystal structure of orthorhombic group IV-VI monolayers $\mathrm{AB}$ (A=Ge and Sn; B=S and Se); (b) the corresponding  Brillouin-zone with the high symmetry points G, X, S and Y. }\label{st}
\end{figure}
\begin{figure*}
  \includegraphics[width=16.0cm]{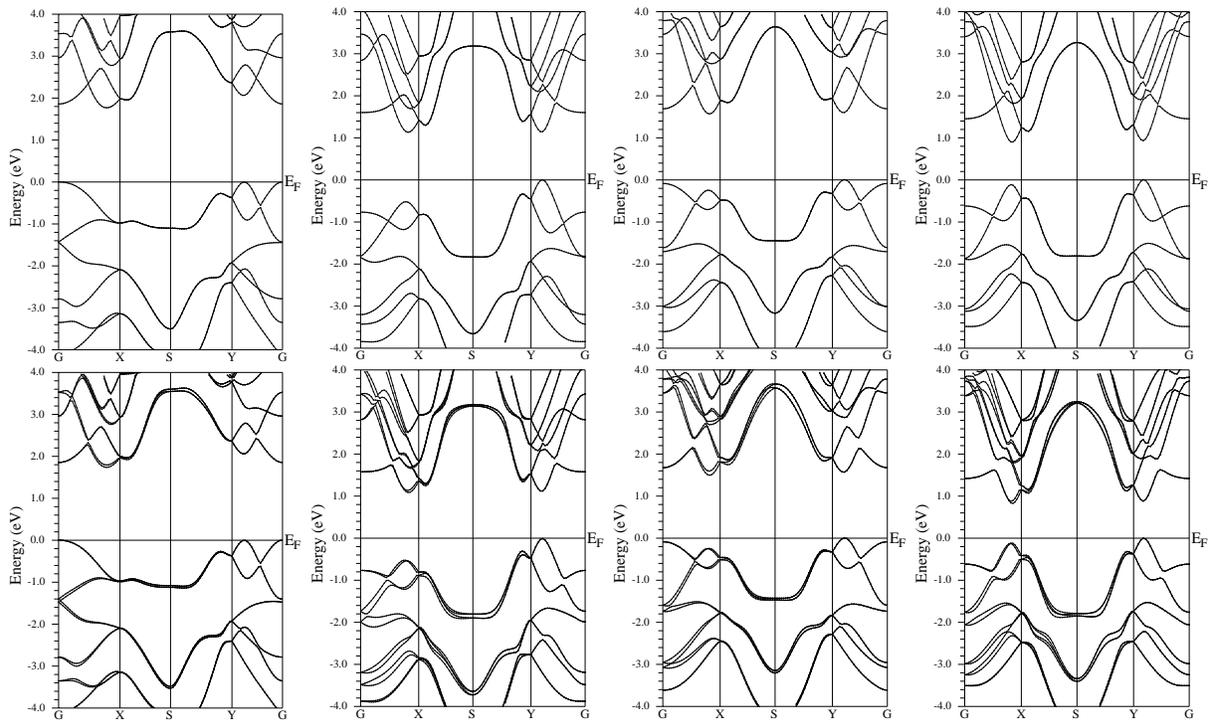}
\caption{The energy band structures of orthorhombic group IV-VI monolayers  using GGA (Top) and GGA+SOC (Bottom). They from left to right are GeS, GeSe, SnS and SnSe, respectively.}\label{band}
\end{figure*}

The bulk orthorhombic group IV-VI compounds $\mathrm{AB}$ (A=Ge and Sn; B=S and Se) with puckered (hinge-like) layered structure are promising candidates for high-efficient thermoelectric materials due to large Seebeck coefficients, high power factors and low thermal conductivities\cite{t1}. Bulk SnSe is especially a robust thermoelectric material with an unprecedented $ZT$ of 2.6 at 973 K along the b axis due to ultralow thermal conductivity\cite{t2,t3}. Like other layered materials, 2D SnSe has been recently synthesized\cite{t4,t5}, which is reported to be a promising 2D  semiconductor\cite{t6}, and the thermoelectric transport has been also investigated\cite{t7}. Besides 2D
SnSe,  the optical and piezoelectric properties of orthorhombic group IV-VI monolayers $\mathrm{AB}$ (A=Ge and Sn; B=S and Se)
have been also studied\cite{t8,t9,t10}. Moreover, it is predicted that orthorhombic group IV-VI monolayers  are multiferroic with coupled ferroelectricity and ferroelasticity, and  GeS and GeSe of them  can maintain their ferroelasticity and ferroelectricity beyond the room temperature\cite{t11}. Recently, the phonon transport properties of orthorhombic group IV-VI monolayers have been systematically investigated by solving the Boltzmann transport equation (BTE) based on first-principles
calculations, and they possess diverse anisotropic property of  lattice thermal conductivity\cite{t12}.
The average lattice thermal conductivities along the zigzag and armchair directions of group IV-VI monolayers are GeS (6.38 $\mathrm{W m^{-1} K^{-1}}$), GeSe (5.23 $\mathrm{W m^{-1} K^{-1}}$), SnS (3.08 $\mathrm{W m^{-1} K^{-1}}$) and  SnSe (2.77 $\mathrm{W m^{-1} K^{-1}}$),   which  suggests that they may be potential 2D thermoelectric materials due to rather low lattice thermal conductivity compared to other 2D materials.
 \begin{table}[!htb]
\centering \caption{The  lattice constants\cite{t12} $a$ and $b$  ($\mathrm{{\AA}}$) along zigzag and armchair directions; the calculated gap values  using GGA $G$ (eV) and GGA+SOC $G_{so}$ (eV); $G$-$G_{so}$ (eV);  spin-orbit splitting $\Delta_{so}$ (eV)  at the CBM. }\label{tab}
  \begin{tabular*}{0.48\textwidth}{@{\extracolsep{\fill}}ccccccc}
  \hline\hline
Name& $a$ & $b$ & $G$& $G_{so}$&$G$-$G_{so}$& $\Delta_{so}$\\\hline\hline
GeS&3.671&4.457&1.767&1.736&0.031&0.057\\\hline
GeSe&3.982&4.269&1.129&1.098&0.031&0.046\\\hline
SnS&4.088&4.265&1.564&1.492&0.072&0.099\\\hline
SnSe&4.294&4.370&0.895&0.829&0.066&0.052\\\hline\hline
\end{tabular*}
\end{table}

Here, we report on the electronic structures and thermoelectric properties of  orthorhombic group IV-VI monolayers $\mathrm{AB}$ (A=Ge and Sn; B=S and Se) from  a combination of  first-principles calculations and semiclassical Boltzmann transport theory.
For electronic part, the SOC  is included  to attain reliable  power factor and electronic thermal conductivity, which has been proved be very important for  electron transport in many thermoelectric materials\cite{s6,s7,so1,so21,so2,so3,so4,so5}.
It is found that SOC can produce observable  influence on  power factor in spite of little SOC effect on electronic structures, and the SOC  not only can reduce power factor but it also can enhance one.
The lattice thermal conductivities from Ref.\cite{t12} and empirical scattering time $\tau$=$10^{-14}$ s are used  to estimate
dimensionless  figure of merit $ZT$, which show that four monolayers, although possessing similar
hinge-like structure, show diverse anisotropic thermoelectric properties. The GeS shows the strongest anisotropy for $ZT$ along the zigzag and armchair directions,  while  SnS and SnSe show neglectful anisotropy.

The rest of the paper is organized as follows. In the next section, we shall
describe computational details for first-principle and transport coefficients calculations. In the third section, we shall present the electronic structures and  thermoelectric properties of  orthorhombic group IV-VI monolayers $\mathrm{AB}$ (A=Ge and Sn; B=S and Se). Finally, we shall give our discussions and conclusion in the fourth
section.

\begin{figure*}
  \includegraphics[width=14cm]{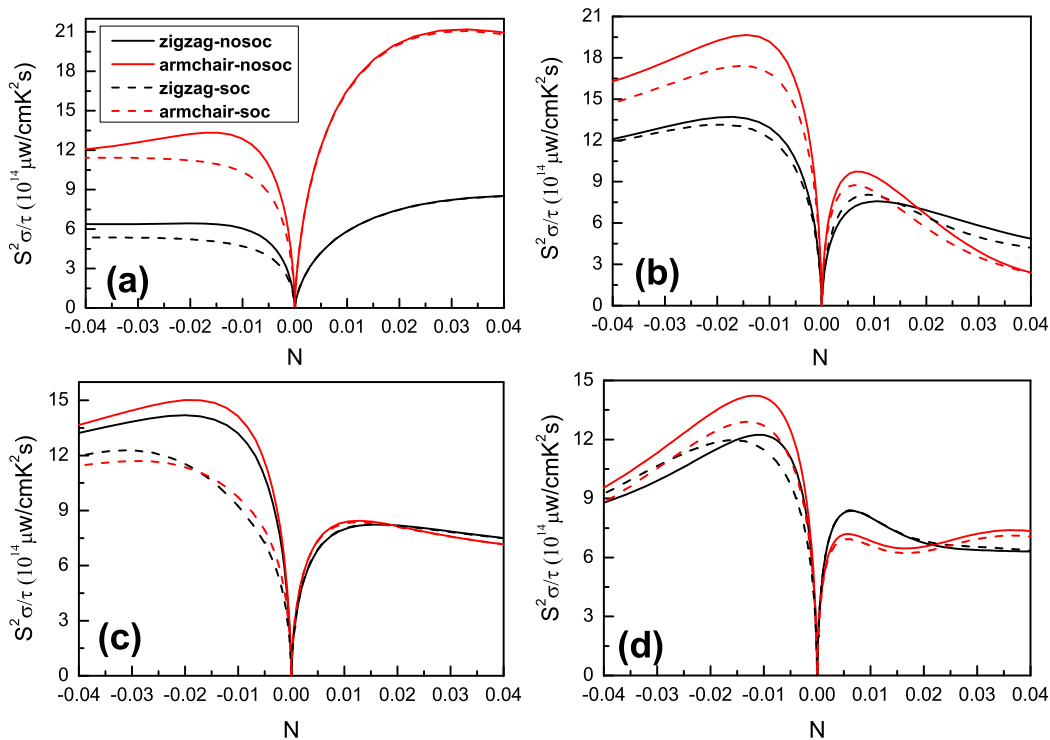}
  \caption{(Color online)  At room temperature (300 K), the power factor with respect to scattering time $\mathrm{S^2\sigma/\tau}$  of GeS (a), GeSe (b), SnS (c) and SnSe (d) for zigzag and armchair directions as a function of doping level (N) using  GGA and GGA+SOC. The doping level (N) implies  electrons (minus value) or holes (positive value) per unit cell.}\label{ft1}
\end{figure*}

\section{Computational detail}
 A full-potential linearized augmented-plane-waves method
within the density functional theory (DFT) \cite{1} is employed to study electronic structures of orthorhombic group IV-VI monolayers $\mathrm{AB}$ (A=Ge and Sn; B=S and Se), as implemented in the WIEN2k  package\cite{2}.  The popular generalized gradient approximation (GGA)\cite{pbe} for the exchange-correlation potential is used to do our electronic structures calculations. The  internal atomic position parameters  are optimized using GGA with a force standard of 2 mRy/a.u..
The SOC was included self-consistently \cite{10,11,12,so} due to containing heavy elements, which leads to band splitting, giving rise to important influences on semi-classic transport coefficients. To attain accurate results, we use at least 6000 k-points in the first Brillouin zone for the self-consistent calculation,  make harmonic expansion up to $\mathrm{l_{max} =10}$ in each of the atomic spheres, and set $\mathrm{R_{mt}*k_{max} = 8}$ . The self-consistent calculations are
considered to be converged when the integration of the absolute
charge-density difference between the input and output electron
density is less than $0.0001|e|$ per formula unit, where $e$ is
the electron charge. Based on the calculated
electronic energy, the  semi-classic transport coefficients, such as  Seebeck coefficient, electrical conductivity and electronic thermal conductivity,
are performed through solving Boltzmann
transport equations within the constant
scattering time approximation (CSTA) as implemented in
BoltzTrap\cite{b} (Note: For 2D materials, the parameter LPFAC usually can not choose the default value 5, and should choose larger value. Here, we choose LPFAC value for 20.),  which  has
been proved to be very effective  for several materials\cite{b1,b2,b3}. To
obtain accurate transport coefficients, at least 22000 k-points  are used  in the
first Brillouin zone for the energy band calculation.
\begin{figure*}
 \includegraphics[width=16cm]{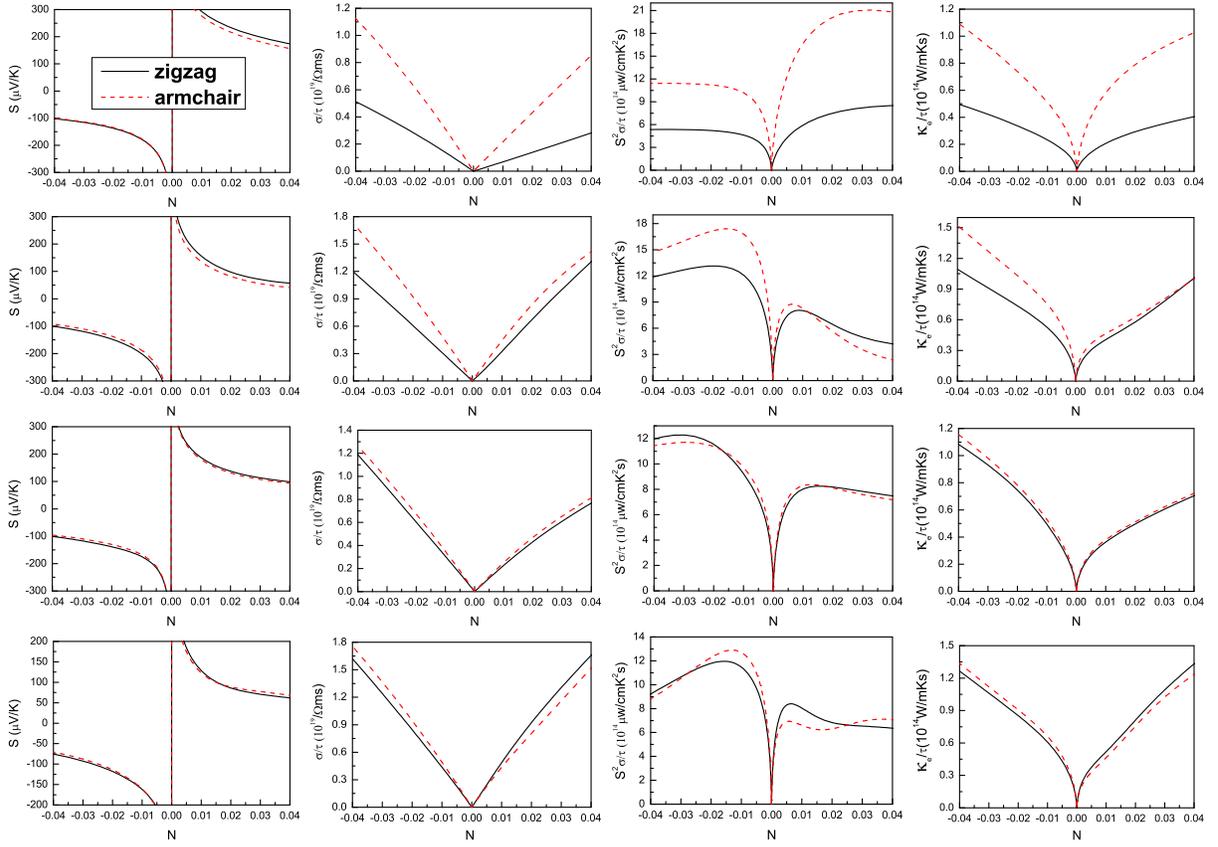}
  \caption{(Color online)  At room temperature,  transport coefficients of GeS (First layer), GeSe (Second layer), SnS (Third layer) and SnSe (Fourth layer) for zigzag and armchair directions as a function of doping level (N):  Seebeck coefficient S, electrical conductivity with respect to scattering time  $\mathrm{\sigma/\tau}$,   power factor with respect to scattering time $\mathrm{S^2\sigma/\tau}$  and electronic thermal conductivity with respect to scattering time $\mathrm{\kappa_e/\tau}$ using GGA+SOC.}\label{fs1}
\end{figure*}

\section{MAIN CALCULATED RESULTS AND ANALYSIS}
The orthorhombic group IV-VI monolayers $\mathrm{AB}$ (A=Ge and Sn; B=S and Se) possess  hinge-like structure, A (B) of which  is covalently bonded to three neighbors of B (A), forming zigzag and armchair directions.  The unit cell  of  monolayer $\mathrm{AB}$ (A=Ge and Sn; B=S and Se) contains  two A and two B atoms, which  is constructed with the vacuum region of more than 15 $\mathrm{{\AA}}$ to avoid spurious interaction. The schematic crystal structure and corresponding Brillouin-zone are shown in \autoref{st}. The space group of monolayers $\mathrm{AB}$ (A=Ge and Sn; B=S and Se) is $Pmn2_1$ (No. 31), possessing lower symmetry than phosphorene  with $Pmna$ (No. 53), which is due to different
types of atoms constituting the compounds compared with phosphorene.  The two sublayers of phosphorene
are  parallel to each other, but not for  group IV-VI monolayers. The optimized  lattice constants $a$ and $b$ along zigzag and armchair directions\cite{t12} are used to investigate their electronic structures and thermoelectric properties, which are summarized  in \autoref{tab}.
\begin{figure*}
  \includegraphics[width=16cm]{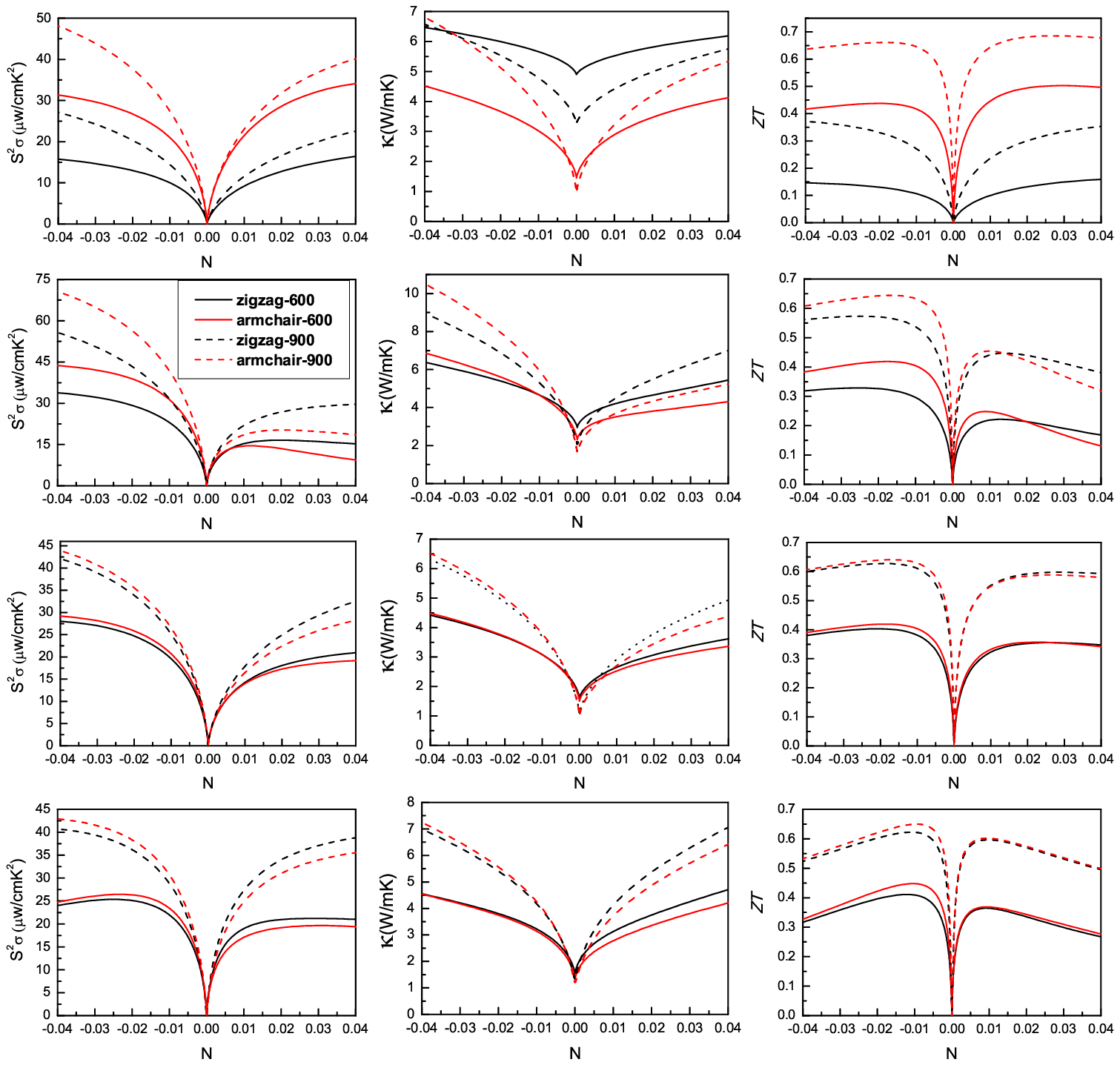}
  \caption{(Color online) At 600 K and 900 K,  the  power factor  $\mathrm{S^2\sigma}$, total thermal conductivity  $\mathrm{\kappa}$  and $ZT$  of GeS (First layer), GeSe (Second layer), SnS (Third layer) and SnSe (Fourth layer) for zigzag and armchair directions as a function of doping level with the scattering time $\mathrm{\tau}$  being 1 $\times$ $10^{-14}$ s. }\label{ft4}
\end{figure*}

Firstly, we investigate the electronic structures of group IV-VI monolayers $\mathrm{AB}$ (A=Ge and Sn; B=S and Se) using GGA and GGA+SOC, and show their energy band structures in \autoref{band}. Both GGA and GGA+SOC results show they all are indirect-gap semiconductor  with the conduction band minimum (CBM) being between the G and X points and  valence band maximum (VBM) being between the Y and G points (Note: For GeS, the VBM is at G point. However, this depends quite sensitively on the lattice constants. In Ref.\cite{t9}, the VBM is still between the Y and G points with a=3.68 $\mathrm{{\AA}}$ and b=4.40 $\mathrm{{\AA}}$.). The related gaps with GGA and GGA+SOC and the differences between them are summarized in \autoref{tab}, which are consistent with previous theoretical results\cite{t9}. The SOC has smaller effects on group IV-VI monolayers than transition-metal dichalcogenide monolayers\cite{s7}, and the representative spin-orbit splitting values   at the CBM  are listed in \autoref{tab}, which are in agreement with previous theoretical values\cite{t9}.
 Among  group IV-VI monolayers, the SOC produces the largest effects on valence bands near the Fermi level for SnS, which will lead to obvious influences on p-type power factor of SnS.
 According to their energy band structures,  it is found that the symmetry along G-X-S and G-Y-S directions gradually increases from GeS to SnSe, which is because  the difference between $a$ along zigzag direction and $b$ along armchair direction  gradually decreases.  The smaller difference leads to less anisotropic  thermoelectric properties. The conduction and valence bands near the Fermi level  also show a certain symmetry, especially for SnS and SnSe.

The SOC has very important influences on electronic transport coefficients in many thermoelectric materials\cite{s6,s7,so1,so21,so2,so3,so4,so5}. Here, we firstly consider SOC effects on transport coefficients of group IV-VI monolayers. On the basis of energy  band structure with GGA and GGA+SOC, the semi-classic transport coefficients, including Seebeck coefficient S and   electrical conductivity with respect to scattering time  $\mathrm{\sigma/\tau}$,  are calculated  within CSTA Boltzmann theory. The rigid band approach is employed, which is  reasonable, if the doping level is low\cite{tt9,tt10,tt11}. The doping effects are mimicked by shifting the Fermi level.
The  power factor with respect to scattering time $\mathrm{S^2\sigma/\tau}$ along zigzag and armchair directions as  a function of doping level (N) at  room temperature  using GGA and GGA+SOC are shown in \autoref{ft1}. Calculated results show that SOC has obvious detrimental effects on power factor in n-type doping for GeS and SnS, while has   negligible influences for p-type. For GeSe and SnSe, the power factor  along armchair direction with GGA+SOC is reduced compared to one with GGA, while slightly improved effect is observed along zigzag direction. These can be understood by considering SOC effects on the conduction or valence bands near the Fermi level. The SOC  can remove the band degeneracy  by  spin-orbit splitting, leading to reduced S, and gives rise to reduced power factor. However, SOC-removed band degeneracy also can make two band extrema to be more close, which can induce improved S, and then enhances  power factor. Similar SOC effects on power factor can also be found in semiconducting transition-metal dichalcogenide monolayers\cite{s6,s7}. For SnS, the maximum power factors along zigzag and armchair directions at 300 K with SOC are predicted to be  about 13.47\%  and 22.07\%   smaller than those  without SOC in the case of n-type doping.
Therefore,  it is necessary to consider SOC effects for theoretical analysis of thermoelectric properties in group IV-VI monolayers.

Next, room-temperature transport coefficients of group IV-VI monolayers $\mathrm{AB}$ (A=Ge and Sn; B=S and Se) for zigzag and armchair directions as a function of doping level, including  Seebeck coefficient S, electrical conductivity with respect to scattering time  $\mathrm{\sigma/\tau}$,   power factor with respect to scattering time $\mathrm{S^2\sigma/\tau}$  and electronic thermal conductivity with respect to scattering time $\mathrm{\kappa_e/\tau}$, are plotted in \autoref{fs1} within GGA+SOC.
In n-type doping, the anisotropy of  thermoelectric transport coefficients along zigzag and
armchair directions is very obvious for GeS and GeSe, while only anisotropy of GeS is very remarkable in p-type doping.
These can be explained by their feature of energy band structure. The profile of energy bands along G-X-S and G-Y-S directions for GeS and GeSe has weaker symmetry than one of SnS and SnSe. It is found that n-type doping has more better power factor than p-type one for GeSe, SnS and SnSe, while it is opposite for GeS. Another notable thing is that the GeS along armchair direction in p-type doping shows highest power factor, which can be explained by band convergence\cite{s1000}. The valence band extrema (VBE) along Y-G and  at G point   are very close,  and the energy difference only is  0.007 eV, which leads to large S, inducing high power factor. The electronic thermal conductivity has similar outlines with electrical conductivity, which is because that
the electrical thermal conductivity is connected with electrical conductivity  by the Wiedemann-Franz law: $\mathrm{\kappa_e}$=
 $L$$\mathrm{\sigma}$$T$, where $L$ is the Lorenz number.

Finally, the figure of merit $ZT$ is calculated to estimate the efficiency of thermoelectric conversion, which needs scattering time  $\mathrm{\tau}$ and lattice thermal conductivity  $\mathrm{\kappa_L}$. It is challenging to  calculate  scattering time $\mathrm{\tau}$  from the first-principle calculations because of the complexity of various carrier scattering mechanisms.
Here, a typical $\tau$=$10^{-14}$ s  is used to attain power factor and electrical thermal conductivity (In Ref.\cite{t1}, this value is also adopted in the thermoelectric calculations of bulk orthorhombic IV-VI compounds.). In Ref.\cite{t12}, the  lattice thermal conductivities of group IV-VI monolayers $\mathrm{AB}$ (A=Ge and Sn; B=S and Se) have been investigated  in detail. The lattice thermal conductivities along the zigzag and armchair directions  show the strongest
anisotropy for GeS,  while the ones of  monolayer SnS and SnSe are very weak, which is similar with electron transport.
The lattice thermal conductivities of four monolayers all almost go as 1/T at medium temperatures. The lattice thermal conductivities of group IV-VI monolayers along zigzag and armchair directions at 600 and 900 K are attained from room-temperature ones. At 600 K and 900 K,  the  power factor  $\mathrm{S^2\sigma}$, total thermal conductivity  $\mathrm{\kappa}$  and $ZT$  of group IV-VI monolayers $\mathrm{AB}$ (A=Ge and Sn; B=S and Se) for zigzag and armchair directions as a function of doping level with the scattering time $\mathrm{\tau}$  being 1 $\times$ $10^{-14}$ s are plotted in \autoref{ft4}. It is found  that the anisotropic behavior
of $ZT$ is keeping pace with that of power factor,  electronic and lattice thermal conductivity.
For  GeS and n-type GeSe, the $ZT$ along  armchair direction  is much larger than one along zigzag direction. The $ZT$ values
along zigzag and armchair directions of p-type GeSe, SnS and  SnSe show less anisotropic behavior, especially for SnS.
Another interesting thing is that n- and p-type $ZT$ values of GeS,  SnS and SnSe show little difference. For GeSe,  n-type doping shows better thermoelectric performance than p-type one. According to their average $ZT$ along zigzag and armchair directions, it  is clearly shown that n-type GeSe, SnS and SnSe exhibit almost excellent thermoelectric performance, while GeS and p-type GeSe have relatively weak one.

\section{Discussions and Conclusion}
Compared to  transition-metal dichalcogenide monolayers, the SOC has rather little effect on electronic structures of orthorhombic group IV-VI monolayers.  The spin-orbit splitting of transition-metal dichalcogenide monolayers at representative point is 0.09 eV$\sim$0.49 eV\cite{s6,s7}, which is larger than that of group IV-VI monolayers from 0.046 eV to 0.099 eV except for  $\mathrm{ZrS_2}$ (0.09 eV).
However, SOC can induce observable influence on power factor of group IV-VI monolayers, especially for n-type SnS due to the largest  spin-orbit splitting at CBM. Both reduced and enhanced effects on power factor induced by SOC are found in group IV-VI monolayers, which is similar with transition-metal dichalcogenide monolayers\cite{s6,s7}. Unlike bulk $\mathrm{Mg_2Sn}$\cite{so21} and half-Heusler $\mathrm{ANiB}$ (A=Ti, Hf, Sc, Y; B=Sn, Sb, Bi)\cite{so2}, only detrimental influences on power factor are observed. So, it is necessary for electron transport of group IV-VI monolayers to include SOC.

The electronic structure of 2D materials is quite sensitive to strain, which can induce  band convergence, further enhancing thermoelectric properties.  Group IV-VI monolayers  have some  VBE and conduction band extrema (CBE) near the Fermi level, and their relative positions of energy level can be tuned by strain. For monolayer SnSe,  Both  CBE  and VBE along G-X  or G-Y and at G point can approach each other by strain\cite{q1}, which can achieve band convergence, improving power factor\cite{s1000}.  In fact, the band convergence can be observed in GeS  with a=3.671 $\mathrm{{\AA}}$ and b=4.457 $\mathrm{{\AA}}$ used in the work, and the VBE along G-Y and at G point almost coincide, exhibiting  high power factor. However, if the lattice constants  a=3.68 $\mathrm{{\AA}}$ and b=4.40 $\mathrm{{\AA}}$ in Ref.\cite{t9} are used, the band convergence  should not occur. Similar strain-induced band convergence also can be found in transition-metal dichalcogenide monolayer  $\mathrm{MoS_2}$\cite{s6}. So, it is possible to tune thermoelectric properties of group IV-VI monolayers by strain.

If we assume that the scattering time is fixed, group IV-VI monolayers have more higher power factor than semiconducting transition-metal dichalcogenide monolayers $\mathrm{MX_2}$ (M=Zr, Hf, Mo, W and Pt; X=S, Se and Te) expect for $\mathrm{PtX_2}$ (X=S, Se and Te) at 300 K, which can be easily observed from \autoref{fs1} and Figure 5 in Ref.\cite{s7}.
They have more lower  lattice thermal conductivities than transition-metal dichalcogenide monolayers like $\mathrm{MS_2}$ and $\mathrm{MSe_2}$ (M=Zr, Hf, Mo and W)\cite{lcccc}. The average lattice thermal conductivities along the zigzag and armchair directions of GeS (6.38), GeSe (5.23), SnS (3.08) and  SnSe (2.77)  all are lower than that of  $\mathrm{MoS_2}$ (103.4), $\mathrm{WS_2}$ (141.9), $\mathrm{MoSe_2}$ (54.21), $\mathrm{WSe_2}$ (52.47),  $\mathrm{ZrS_2}$ (13.31),  $\mathrm{HfSe_2}$ (11.30),
$\mathrm{HfS_2}$ (16.56)  and $\mathrm{ZrSe_2}$ (10.10) [The unit of thermal conductivity: $\mathrm{W m^{-1} K^{-1}}$].
Therefore, group IV-VI monolayers may be potential 2D thermoelectric materials.

In summary,  the first-principles
combined with the Boltzmann transport theory are used to investigate the
thermoelectric properties of orthorhombic group IV-VI monolayers $\mathrm{AB}$ (A=Ge and Sn; B=S and Se), and the SOC is also included for electron transport. Although the  SOC influences on electronic structures are not very obvious, SOC-induced splitting produces observable effects on power factor.
The four monolayers show diverse anisotropic thermoelectric  properties,
and GeS  along the zigzag and armchair directions
shows the strongest anisotropy while  SnS and SnSe show an almost isotropy.
In n-type doping, it is found that four monolayers show similar efficiency of thermoelectric conversion along armchair direction,
and GeS shows the weakest one along zigzag direction compared with GeSn, SnS and SnSe.
For p-type, GeSe shows the lowest $ZT$  along armchair direction, and GeS and GeSe exhibit weaker conversion efficiency along zigzag direction. The present work will further stimulate  experimental  studies of 2D high-efficient thermoelectric materials.

\begin{acknowledgments}
This work is  supported by the Fundamental Research Funds for the Central Universities (2015XKMS073). We are grateful to the Advanced Analysis and Computation Center of CUMT for the award of CPU hours to accomplish this work.
\end{acknowledgments}

\end{document}